\documentclass[12pt]{article}
\usepackage{url}
\usepackage{epsfig,epsf}
\usepackage{graphicx,graphics}
\usepackage{amssymb,amsmath}
\usepackage{verbatim,enumerate}
\usepackage{verbatim,multicol,color}
\usepackage{multirow}
\usepackage{booktabs}
\usepackage{multicol}
\usepackage[english]{babel}
\usepackage{blindtext}
\usepackage{subfig}
\usepackage{natbib}
\usepackage{float}

\usepackage{color}
\usepackage[normalem]{ulem}

\def\log{\hbox{log}}

\def\boxit#1{\vbox{\hrule\hbox{\vrule\kern6pt
          \vbox{\kern6pt#1\kern6pt}\kern6pt\vrule}\hrule}}

\def\bse{\begin{eqnarray*}}
\def\ese{\end{eqnarray*}}
\def\be{\begin{eqnarray}}
\def\ee{\end{eqnarray}}
\def\bq{\begin{equation}}
\def\eq{\end{equation}}
\def\bse{\begin{eqnarray*}}
\def\ese{\end{eqnarray*}}

\usepackage{bm}

\newcommand{\bsy}{\boldsymbol}
\newcommand{\iid}{\stackrel{\mathrm{iid}}{\sim}}

\newcommand{\mbf}{\mathbf}
\newtheorem{thm}{Result}

\begin{document}
\thispagestyle{empty}
\baselineskip=28pt
\vskip 5mm
\begin{center} {\Large{\bf An Evolutionary Spectrum Approach to Incorporate Large-scale Geographical Descriptors on Global Processes}}
\end{center}

\baselineskip=12pt
\vskip 5mm

\begin{center}
\large
Stefano Castruccio{\footnotemark[1]} and Joseph Guinness{\footnotemark[2]}  
\end{center}

\footnotetext[1]{
\baselineskip=10pt School of Mathematics \& Statistics, Newcastle University, Newcastle Upon Tyne, NE1 7RU United Kingdom. E-mail: stefano.castruccio@ncl.ac.uk}

\footnotetext[2]{
\baselineskip=10pt Department of Statistics, North Carolina State University, 2311 Stinson Drive, Raleigh, NC 27695, United States. E-mail: joeguinness@ncsu.edu}
\baselineskip=16pt
\vskip 4mm
\centerline{\today}
\vskip 6mm

\begin{center}
{\large{\bf Abstract}}
\end{center}

We introduce a nonstationary spatio-temporal model for gridded data on the sphere. The model specifies a computationally convenient covariance structure that depends on heterogeneous geography. Widely used statistical models on a spherical domain are nonstationary for different latitudes, but stationary at the same latitude (\textit{axial symmetry}). This assumption has been acknowledged to be too restrictive for quantities such as surface temperature, whose statistical behavior is influenced by large scale geographical descriptors such as land and ocean. We propose an evolutionary spectrum approach that is able to account for different regimes across the Earth's geography, and results in a more general and flexible class of models that vastly outperforms axially symmetric models and captures longitudinal patterns that would otherwise be assumed constant. The model can be estimated with a multi-step conditional likelihood approximation that preserves the nonstationary features while allowing for easily distributed computations: we show how the model can be fit to more than 20 million data points in less than one day on a state-of-the-art workstation. The resulting estimates from the statistical model can be regarded as a synthetic description (i.e. a compression) of the space-time characteristics of an entire initial condition ensemble.

\baselineskip=16pt

\par\vfill\noindent

\par\medskip\noindent
{\bf Short title}: land/ocean nonstationarity

\clearpage\pagebreak\newpage \pagenumbering{arabic}
\baselineskip=26pt

\section{Introduction}

Providing efficient and flexible models for data on a spherical domain is a topic of great importance in climate science, as the statistical model can be used to fit global data. In particular, in the context of Earth System Models (ESMs), this could lead to efficient methods for compressing large quantities of data. Isotropic models have been widely acknowledged as being inadequate for data on a spherical domain \citep{gn13}, and defining valid nonstationary processes is listed among the sixteen open problems in modeling spherical data in \cite{gn13s}. By regarding a random field as solution of a stochastic partial differential equation \citep{li11} on a spherical domain, \cite{bo11} proposed a nested stochastic partial differential equations approach, which yielded a field with Mat\'ern-like covariance structure but could also be extended beyond axial symmetry to nonstationary models by allowing flexible differential operators in the stochastic equation. \cite{ju07,ju08a,ju08b,ju11} restrict three-dimensional isotropic fields to a sphere and apply partial derivatives with respect to latitude and longitude,  obtaining a model which assumes stationarity if the data are at the same latitude, and nonstationarity otherwise (\textit{axially symmetric} \citep{jo63}, see theoretical details in \cite{hi11,hu12}). Such models are conceptually attractive for data such as surface temperature, whose statistical properties clearly depend on latitude. \cite{ca13b} and \cite{ca14} proposed a spectral approach that is flexible and computationally efficient when the data are on a regular grid over the sphere. This method proposes to separately consider the process by latitudinal bands, fit a Mat\'ern-like covariance across longitudes, and then estimate the multi-band dependence, thus reducing the likelihood evaluation with the full dataset to a low dimensional parameter space. The main limitation of these models, as acknowledged in the aforementioned literature, is the assumption of stationarity in longitude at each latitude. 

For physical quantities such as surface temperature, it is expected that large scale geographical descriptors such as land/ocean will impact the statistical behavior of the data. Recently \cite{ju14} proposed a modified Mat\'ern process with smoothness changing over land and ocean, which showed dramatic improvements over the axially symmetric model. The model parameters, however, were not simple to interpret given their definition through a differential operator over an isotropic process and the fitting procedure was not computationally feasible for analyzing millions of data points. 

This work introduces a new class of covariance functions on spheres that includes axially symmetric models as special cases and is capable of incorporating geographic covariates into the model. For the surface temperature data we consider, the most prominent and influential large scale geographic descriptor is land versus ocean, so we focus our work on this covariate, but as we describe in Section \ref{mod_tot}, the ideas generalize to other covariates as well. We also propose a reformulation of the latitudinal dependence of the model in terms of a stationary AR(1) process, and introduce a nonstationary generalization which is more flexible in capturing different behaviors in the tropics. 

For inference, we devise a step-wise conditional likelihood approach that fully exploits the gridded geometry of climate model output and is able to achieve a fit of more than 20 million data points in less than one day by allowing code parallelization on a state-of-the-art workstation. The proposed method vastly outperforms the axially symmetric model in terms of standard model selection metrics, and is also able to capture patterns in the longitudinal contrasts that would be otherwise assumed constant. The set of estimated parameters can then be used to almost instantaneously produce new surrogate simulations on a common laptop, thus allowing an end user to conveniently test initial scientific hypotheses on a high (spatial) resolution ensemble without downloading it, or remotely aggregating data in space or time and losing valuable information at fine scale. Besides, the estimated parameters can be regarded as descriptors of the information for every member of the given initial ensemble \citep{ca16}, and thus as a compression algorithm \citep{ri89,ha01}. The proposed statistical model achieves a compression rate of approximately 3:100, which is vastly superior to traditional bit-by-bit compression algorithms that can achieve at most a 1:5 ratio.

The model can also be viewed as an emulator of an initial condition ensemble \citep{ca16}, under the assumption that runs are independent for different initial conditions \citep{lo63,co02a,co02b,br10}. The use of emulators as data compressors is, to our knowledge, new to the climate community as they are traditionally used for calibration and sensitivity analysis \citep{sa08,sa09,bh12,dr08,ch13} or scenario extrapolation \citep{ho10,ho13,ca14a}. Having statistical models (emulators) that accurately describe the model output allows us to avoid storing the entire initial condition ensemble, whose individual member requires significant storage space. This proposed methodology has shown promising results and can be generalized to multiple variables (not necessarily in the atmospheric part of the model), climate models and scenarios, and to finer temporal scales. In all these cases, the benefits of a statistical-based data compression will be even more evident as the size of the data, and consequently the expected time of downloading the full climate run, will significantly increase.

The remainder of the paper is organized as follows. Section \ref{sec_data} introduces the data set and discusses nonstationarity across longitude. Section \ref{sec_model} describes the statistical model, discusses the computational challenges that arise when fitting this with very large data sets and suggests a stepwise model-fitting approach to address these challenges in a way that exploits the geometry of the sphere. Section \ref{sec_comparison} shows the comparison with the axially symmetric model. Section \ref{sec_simulation} shows how the fitted model can be used to compress the initial condition ensemble and how to generate surrogate runs from the estimated parameters. Section \ref{sec_conclusion} concludes with a discussion.

\section{The CMIP5-CCSM4 ensemble}\label{sec_data}

\begin{sloppypar}
The Coupled Model Intercomparison Project phase 5 \citep[CMIP5][]{ta12} is a set of coordinated experiments from many modeling groups to provide uniform and comparable assessment of climate response under different climate models for the latest IPCC Assessment Report \citep{ip13}. In this work we focus on the NCAR Community Climate System Model 4 \citep[CCSM4][]{ge11}, under the Representative Concentration Pathway 8.5 \citep[rcp85][]{vv11} from 2006 to 2100, for a total of 95 years. We consider annual temperature at surface (considered at a standard height of 2 meters above ground level), which is on a regular $192\times 288$ grid over latitude and longitude. We remove the bands near the poles (south of 62 degrees south and north of 70 degrees north) so that each spatial field consists of $142\times 288$ points. The removal of the Arctic and Antarctic bands was performed to avoid inference on latitudes where two locations at one longitudinal lag were very close, consistently with \cite{ca13b,ca14} and \cite{ca16}. This would have led to very smooth spatial processes, and consequently computational challenges that would have added to the already substantial complexity of the inference scheme. Under rcp85, the CCSM4 was run under 6 different sets of initial conditions, therefore generating 6 independent realizations \citep{lo63,co02a,co02b,br10}. The total size of the dataset is therefore $142\times288\times95\times6=23.3$ million points. The movie movie\_cm.avi in the supplementary material shows a realization of this climate model. The annual temperature shows evidence of normality, as reported in the supplementary material.
\end{sloppypar}
We denote by $\mathbf{T}_r$ the temperature for realization $r=1,\ldots,R$, by $L_m\in (-\pi/2,\pi/2), m=1,\ldots,M$ the latitude, by $\ell_n=2\pi n/N, n=1,\ldots, N$ the longitude, by $t_k, k=1,\ldots,K$ the year, where $R=6$, $M=142, N=288$ and $K=95$. Thus, the temperature for realization $r$ is represented as
\[
\mbf{T}_r=\{\mbf{T}_r(L_1,\ell_1,t_1),\ldots,\mbf{T}_r(L_M,\ell_1,t_1),\mbf{T}_r(L_1,\ell_2,t_1),\ldots,\mbf{T}_r(L_M,\ell_N,t_K)\}.
\]

\begin{figure}[ht]
\hspace{-0.5em}
\centerline{\includegraphics[width=16cm,keepaspectratio]{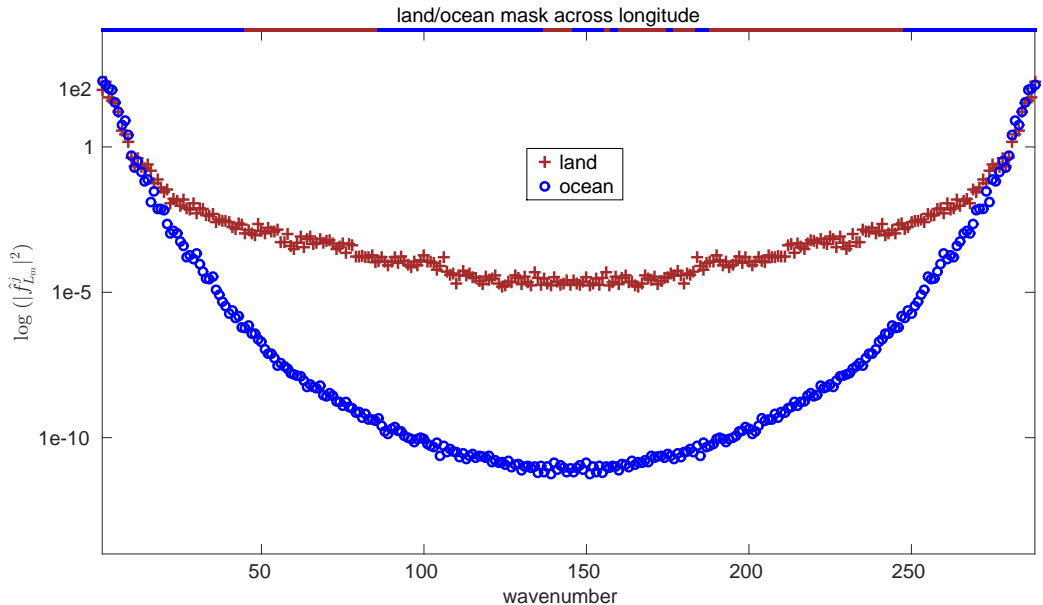}}
\caption{Comparison of the land/ocean periodogram of the difference between two realizations of rcp85 computed with \eqref{perio_ev}, each pixel being normalized by its standard deviation from the control run. The latitude band represented is $\approx 41^\circ$N, and the periodogram is averaged over all years. On the top the land/ocean mask across longitude is represented.}
\label{85_norm_per}
\end{figure}

The defining assumption of axially symmetric models is that the process is stationary across longitude at each latitude. In this work, we relax this assumption to allow geographic covariates to be incorporated into the covariance function and inform more complex spatial dependence structures. Local geography can have a strong impact on the statistical characteristics of surface temperature data, so a natural deviation from the stationary assumption is to allow the statistical properties of the process to differ over land and ocean, which is the most dramatic geographic descriptor at large scales. A simple modeling solution is to divide the temperature at each location $(L_m,\ell_n)$ by the standard deviation $s_{L_m,\ell_n}$ obtained from a simulation from the same climate model but with no forcing (a \textit{control run} in geoscience terminology), as proposed in \cite{ca13b}, to obtain more realistic conditional simulations. While producing improved results, this approach does not allow for a changing correlation structure across longitude, and in particular across land and ocean. Indeed, empirical estimates of the second-order (covariance) structure show a strong dependence on the land/ocean variable. To see this, we consider the difference between two realizations $\mbf{T}_1-\mbf{T}_{2}$ (to remove any trend), normalize it by $s_{L_m,\ell_n}$, and compute 
\begin{equation}\label{perio_ev}
|\hat{f}_{L_m}^j(c)|^2 = \frac{1}{K}\sum_{k=1}^K\frac{p(h^j)}{N} \left|\sum_{n = 1}^N h^j(\ell_n)\frac{\mbf{T}_1(L_m,\ell_n,t_k)-\mbf{T}_{2}(L_m,\ell_n,t_k)}{s_{L_m,\ell_n}} e^{-i\ell_n c}\right|^2,
\end{equation}
for $j=1,2$, which is the periodogram of a tapered version of the data averaged over time at latitude $L_m$, where the taper $h^1$ is a smooth function that is equal to zero when $\ell_n \in$ ocean, and $h^2$ is a smooth function that is equal to zero when $\ell_n \in$ land, and $p(h^j)$ is a normalizing constant. Thus we can view $|\hat{f}_{L_m}^1|^2$ as a periodogram for the land data averaged over time at latitude $L_m$, and $|\hat{f}_{L_m}^2|^2$ as a periodogram for the ocean data averaged over time at latitude $L_m$. In Figure \ref{85_norm_per}, we plot $\log(|\hat{f}_{L_m}^j|^2)$ at latitude $L_m=41^{\circ}$. Because the two log periodograms in Figure \ref{85_norm_per} are not parallel--which would indicate similar correlation structure over land and ocean--it is clear that the data exhibit land/ocean nonstationary correlation, and that the process over the ocean is much smoother than the process over the land at this latitude.

The land and ocean periodograms in Figure \ref{85_norm_per} motivate the use of a model that allows for a different behavior across land and ocean for the global space-time temperature data. In this work, we define a nonstationary model with changing spectrum across these two domains, denoted as \textit{evolutionary spectrum}. The details are provided in Section \ref{mod_tot}

\section{The space time model} \label{sec_model}
In this section we describe the global space-time model. We first introduce in Section \ref{prel} a fundamental result that allows us to fit the stochastic component of the statistical model without defining a parametric form for the mean. In Section \ref{mod_tot} we describe the model. Section \ref{mod_comp} shows the results and discusses the computational challenges of fitting the proposed model on a dataset with tens of millions of data points. 

\subsection{Preliminaries}\label{prel}
Denote by $\mathbb{E}(\mathbf{T}_r)=\boldsymbol{\mu}$ the mean temperature across realizations. Since realizations differ just in their initial condition, and since climate models tend to forget their initial state after a short number of temporal steps \citep{lo63,co02a,co02b,br10}, we can assume that the space-time field $\mbf{T}_r$ is independent across $r$:
\begin{equation}\label{mod1}
\mbf{T}_r=\boldsymbol{\mu}+\boldsymbol{\varepsilon}_r, \qquad
\boldsymbol{\varepsilon}_r\iid \mathcal{N}(\mathbf{0},\bsy{\Sigma}(\boldsymbol{\theta})), 
\end{equation}
where $\boldsymbol{\theta}$ is a vector of unknown covariance parameters. The noticeable advantage of having independent realizations is that $\boldsymbol{\theta}$ can be estimated without any parametrization of $\bsy{\mu}$ via restricted loglikelihood. \cite{ca13b} proved the following result, which formulates the restricted loglikelihood for $\mbf{T}_r$ in a computationally convenient form.
\begin{thm}\label{remlth}
Denote with $\frac{1}{R}\sum_{r=1}^R \mbf{T}_r=\mbf{\bar{T}}$ the average temperature across realizations. Let $\mbf{D}_r=\mbf{T}_r-\mbf{\bar{T}}$. The negative restricted loglikelihood for \eqref{mod1} is
\begin{equation}\label{reml_me}
\begin{array}{lll}
l(\bsy{\theta};\mbf{D}) & = & \frac{KNM(R-1)}{2}\log(2\pi)+\frac{1}{2}(R-1)\log \left|\Sigma(\bsy{\theta})\right|\\[6pt]
                  &   & +\frac{1}{2}KNM\log(R)-\frac{1}{2}\sum_{r=1}^R \mbf{D}_r^{\top} \Sigma(\boldsymbol{\theta})^{-1} \mbf{D}_r.
\end{array}
\end{equation}
Also, the corresponding estimator for $\bsy{\mu}$ obtained by generalized least squares is $\hat{\bsy{\mu}}=\mbf{\bar{T}}$.
\end{thm}
Throughout this work we make use of \eqref{reml_me} to estimate the space/time structure of the data.

\subsection{Sphere-Time Covariance}\label{mod_tot}
\begin{sloppypar}
Denote by $\bsy{\varepsilon}(t_k;r)=\{\bsy{\varepsilon}_r(L_1,\ell_1,t_k),\ldots,\bsy{\varepsilon}_r(L_N,\ell_M,t_k)\}$ the vector of the stochastic term of \eqref{mod1} at time $t_k$, by $\mbf{T}(t_k;r)$ the temperature at year $t_k$ for realization $r$ and by $\mbf{D}(t_k;r)=\mbf{T}(t_k;r)-\mbf{\bar{T}}(t_k;r)$. We assume that $\bsy{\varepsilon}(t_k;r)$ is correlated across time, and previous work \citep{ca14} has shown that an AR(2) model with different coefficients for every grid point is sufficiently flexible, as no evidence of cross-temporal dependence or nonseparability between space and time was found on annual scale:
\end{sloppypar}
\begin{equation}
\bsy{\varepsilon}(t_k;r) = \bsy{\Phi}_1 \bsy{\varepsilon}(t_k-1;r)+\bsy{\Phi}_2 \bsy{\varepsilon}(t_k-2;r)+\mbf{S}\mbf{H}(t_k;r), 
\label{AR1}
\end{equation}
where $\bsy{\Phi}_1$ and $\bsy{\Phi}_2$ are two $NM\times NM$ diagonal matrices with the autoregressive coefficients for each location, and $\mbf{S}$ is a $NM\times NM$ diagonal matrix with the associated standard deviations. 

The unscaled innovations $\mbf{H}_r(L_m,\ell_n,t_k)$ are independent across time and describe the spatial dependence across the sphere. We propose to model the process in the spectral domain across longitudes, and then to model the dependence across latitudes:
\begin{equation}\label{lo_reg}
\begin{array}{l}
\mbf{H}_r(L_m,\ell_n,t_k) =  \sum_{c=0}^{N-1} f_{L_m,\ell_n}(c)e^{i\ell_n c}\widetilde{\mbf{H}}_r(c,L_m,t_k),\\
\mbox{corr}\left\{\widetilde{\mbf{H}}_r(c,L_m,t_k),\widetilde{\mbf{H}}_{r'} (c',L_{m'},t_{k'})\right\} = \mbf{1}\{c=c',k=k',r=r'\}\rho_{L_m,L_{m'}}(c),
\end{array}
\end{equation}
where $c$ is a wavenumber. If $f_{L_m,\ell_n}=f_{L_m}$ for every $L_m$, than this would be a standard stationary model with spectral density $|f_{L_m}|^2$. This proposed model assumes that the spectral density is not exactly constant in longitude, but evolves (hence the term \textit{evolutionary spectrum}) according to $|f_{L_m,\ell_n}|^2$. Models with evolutionary spectra allow us to flexibly specify the local covariance properties at every location while ensuring that the resulting covariance function is positive definite. Evolutionary spectra were first introduced by \cite{priestley1965evolutionary} to model nonstationary time series data, and \citet{gu13a} provided computationally efficient methods for fitting models with evolutionary spectra to nonstationary time series data. In this work, we adapt the evolutionary spectra to model nonstationarity over longitude rather than over time, which requires a discrete spectrum because of the circular domain on which the data at a single latitude fall. 

Certain regularity conditions can be imposed on $f_{L_m,\ell_n}$ near the poles to achieve mean square continuity (see supplementary material). $\rho_{L_m,L_{m'}}$ is the correlation in the spectral domain (or \textit{coherence} in spectral analysis) between latitudes $L_m$ and $L_{m'}$ among $\widetilde{\mbf{H}}_r$ for the same wavenumber, time and realization. Alternatively, one could deviate from the stationary assumption across longitude by introducing dependence across wavenumber in $\widetilde{\mbf{H}}_r$. Our choice to use evolutionary spectra to model the nonstationarities is motivated by the need to include geographic covariates.

The typical approach \citep{priestley1965evolutionary} for $f_{L_m,\ell_n}(c)$ is to describe the dependence on $\ell_n$ according to covariates $X^j(L_m,\ell_n)$ as
\begin{equation}
f_{L_m,\ell_n}(c) = \sum_{j = 1}^p f^j_{L_m}(c) X^j(L_m,\ell_n).
\end{equation}
In this work, we propose a novel model where land and ocean are included as covariates to allow for different statistical behaviors across these two domains. Thus, $f_{L_m,\ell_n}(c)$ can be expressed as
\begin{equation}\label{quax_eq1}
f_{L_m, \ell_n}(c)=f^{1}_{L_m}(c)b_{\mbox{land}}(L_m,\ell_n)+f^{2}_{L_m}(c)\left\{1-b_{\mbox{land}}(L_m,\ell_n)\right\},
\end{equation}
where the component spectra are modeled according to the parametric form
\begin{equation}\label{quax_eq2}
|f^{j}_{L_m}(c)|^2=\frac{\phi^j_{L_m}}{\left\{(\alpha^{j}_{L_m})^2+4\sin^2\left(\frac{c}{N}\pi\right)\right\}^{\nu^j_{L_m}+1/2}}, \qquad j=1,2,  
\end{equation}
and $b_{\mbox{land}}$ is a function between 0 and 1 that modulates the relative contribution of the land regime. \eqref{quax_eq2} is a Mat\'ern-like spectrum, which is modified for the case of data on a circle to allow for a smooth transition at high wavenumbers and has been shown to adequately capture the longitudinal behavior of temperature at surface for different latitudes better than the traditional Mat\'ern-like spectrum \citep{ca13b,po14}.

Choosing an indicator function for $b_{\mbox{land}}$ would result in abrupt transitions between the two regimes at land/ocean boundaries and in misfit of the data, as shown in the supplementary material. We therefore introduce a smoother taper function to transition between land and ocean:
\begin{itemize}
\item Let $I_m(\ell_n)$ denote the indicator function of land at latitude $L_m$ and longitude $\ell_n$. Wherever there is a land/ocean transition, we modify $I_m(\ell_n)$ so that is equal to one for $g_{L_m}$ more grid points, where $g_{L_m}$ is an integer number that can also be negative. The modified indicator is denoted by $\tilde{I}_m(\ell_n;g_{L_m})$. 
\item Compute the Tukey taper function \citep{tu67} with range $\gamma_{L_m}$:
\begin{equation}
w_m(\ell_n;\gamma_{L_m})=\left\{\begin{array}{ll}\frac{1}{2}\left[1+\cos\left\{\frac{2\pi}{\gamma_{L_m}}(\ell_n-\gamma_{L_m}/2) \right\}\right], & 0 \leq \gamma_{L_m} < \frac{\gamma_{L_m}}{2}, \\[7pt] 1, & \gamma_{L_m}/2 \leq \ell_n < 1-\gamma_{L_m}/2, \\[7pt] \frac{1}{2}\left[1+\cos\left\{\frac{2\pi}{\gamma_{L_m}}(\ell_n-1-\gamma_{L_m}/2) \right\}\right], & 1-\gamma_{L_m}/2\leq \ell_n \leq 2\pi.\end{array} \right.
\end{equation}
\item Convolve $\tilde{I}_m(\ell_n;g_{L_m})$ with $w_m(\ell_n;\gamma_{L_m})$:
\begin{equation}\label{qax_conv}
b_{\mbox{land}}(L_m,\ell_n;g_{L_m},\gamma_{L_m})=\sum_{n'=1}^{N}\tilde{I}_m(\ell_n;g_{L_m}) w_m(\ell_n-\ell_{n'};\gamma_{L_m}).
\end{equation}
\end{itemize}
This formulation imposes a symmetric land/ocean transition (i.e. land/ocean and ocean/land transitions are equally smooth); however, more sophisticated models with asymmetric transitions have been tested but have not yielded significantly better results. Similarly, no significant improvements have been observed if a different taper is used (as shown in the supplementary material) or $g_{L_m}$ and $\gamma_{L_m}$ are assumed different across oceans. If we constrain 
\begin{equation}\label{bandstat}
\phi_{L_m}^1=\phi_{L_m}^2, \alpha_{L_m}^1=\alpha_{L_m}^2, \nu_{L_m}^1=\nu_{L_m}^2,
\end{equation}
then in \eqref{lo_reg} $f_{L_m,\ell_n}=f_{L_m}$ and the model becomes stationary across longitude. 

\cite{ca13b,ca14,ca16} propose the following parametric model for $\rho_{L_m,L_{m'}}(c)$ in \eqref{lo_reg}:
\begin{equation}\label{coh}
\rho_{L_m,L_{m'}}(c)=\rho_{L_m-L_{m'}}(c)=\left[\frac{\xi}{\left\{1+4\sin^2\left(\frac{c}{N}\pi\right)\right\}^{\tau}} \right]^{|L_m-L_{m'}|}=\varphi(c)^{|L_m-L_{m'}|}.
\end{equation}
with $\varphi(c)=\frac{\xi}{\left\{1+4\sin^2\left(\frac{c}{N}\pi\right)\right\}^{\tau}}$. This process is equivalent to the following AR(1) process in latitude:
\begin{equation}\label{ar_lat}
\begin{array}{lll}\widetilde{\mbf{H}}_{L_m}(c) & = & \left\{\begin{array}{lll}\varphi(c)\widetilde{\mbf{H}}_{L_{m-1}}(c)+\mbf{e}_{L_m}(c), & m=2,\ldots,M,\\[7pt] \mbf{e}_{L_1}(c) \sim \mathcal{N}(0,1), & m=1, \end{array}\right.\\[18pt]
\mbf{e}_{L_m} & \iid & \mathcal{N}(0,1-\varphi(c)^2), \qquad m>1,
\end{array}
\end{equation}
where $\text{var}(\mbf{e}_{L_m}(c))=1-\varphi(c)^2$ to allow unit variance on $\widetilde{\mbf{H}}_{L_m}(c)$. While the coherence in \eqref{coh} has been previously used in literature, the formulation of the latitudinal dependence in terms of an autoregressive process has never been acknowledged. 

The formulation of the dependence as a stationary AR(1) process allows for generalization to nonstationary latitudinal processes to increase the model flexibility. In particular, in addition \eqref{coh} we also propose a novel nonstationary AR(1) model for the coherences, with latitudinally-varying autoregressive parameters, that is
\begin{equation}\label{ar_nstat}
\varphi_{L_m}(c)=\frac{\xi_{L_m}}{\left\{1+4\sin^2\left(\frac{c}{N}\pi\right)\right\}^{\tau_{L_m}}}.
\end{equation}
Our diagnostics have shown that the coherences are nearly stationary outside of the tropics, so we fit nonstationary coherences within $-23^\circ<L<23^\circ$ (i.e. in the tropics), while we assume a constant outside this region. 

Thus, the model consists of three sets of parameters to be estimated
\begin{itemize}
\item The temporal parameters, consisting of all the entries in $\bsy{\Phi}_1$, $\bsy{\Phi}_2$ and $\mbf{S}$ in \eqref{AR1}, which will be denoted as $\bsy{\theta}_{\text{time}}$.
\item The longitudinal parameters, consisting of $(\phi^{j}_{L_m},\alpha^{j}_{L_m},\nu^{j}_{L_m})$ in \eqref{quax_eq2} and $(g_{L_m},\gamma_{L_m})$ in \eqref{qax_conv} for $m=1,\ldots, M$. We denote the collection of all parameters as $\bsy{\theta}_{\text{lon}}$.
\item The latitudinal parameters, consisting of $(\xi_{L_m},\tau_{L_m})$ in \eqref{ar_nstat} $m=1,\ldots, M$. We denote them as $\bsy{\theta}_{\text{lat}}$.
\end{itemize}

\subsection{Model fit and computational considerations}\label{mod_comp}

Despite the computationally convenient form in \eqref{reml_me} (which can be further simplified as shown in the supplementary material) it is not feasible to perform a global optimization, since this would imply maximizing the likelihood over more than 100,000 parameters  (the temporal part requires 3 parameters for each of the $142\times 288\approx 41,000$ locations, the spatial part requires a total number of parameters shown in the first row of Table \ref{fitres}). We therefore propose successive conditional approximations of \eqref{reml_me} by assuming independence across increasingly large subsets, each approximation assuming the parameters from previous steps to be known and fixed. 
\begin{enumerate}
\item Estimate the temporally autoregressive parameters $\bsy{\theta}_{\text{time}}$, assuming that the innovations $\mbf{H}(t_k;r)$ are independent across latitude and longitude. 
\begin{sloppypar}
\item Consider $\bsy{\theta}_{\text{time}}$ fixed at their estimated values and estimate $\bsy{\theta}_{\text{lon}}$, assuming the innovations $\mbf{H}(t_k;r)$ are independent across latitudes.
\end{sloppypar}
\item Consider $\bsy{\theta}_{\text{time}}$ and $\bsy{\theta}_{\text{lon}}$ fixed at their estimated values and estimate $\bsy{\theta}_{\text{lat}}$.
\end{enumerate}

The choice of the blocks in the approximation, as well as the approximation order is dictated by the geometry of the problem as well as from physical considerations. Estimating the temporal structure for each location assuming no cross-correlation allows for a very fast (and scalable) computation of approximation 1. The choice of latitudinal bands in approximation 2 allows flexible estimation of the statistical parameters across latitude, which is the main descriptor of surface temperature. Further, this choice results in exactly circulant matrices across longitudes in the axially symmetric case, a feature that allows very fast computations in the spectral domain. This conditional approximation scheme can be generalized to allow for vertical profile of temperatures as in \cite{ca16}, and can also be applied to any large space-time data set where the geometry, as well as the physics of the problem suggest the blocks, e.g. regions of interest in functional Magnetic Resonance Imaging \citep{ca16b}.

\begin{figure}[ht]
\hspace{-0.5em}
\centerline{\includegraphics[width=16cm,keepaspectratio]{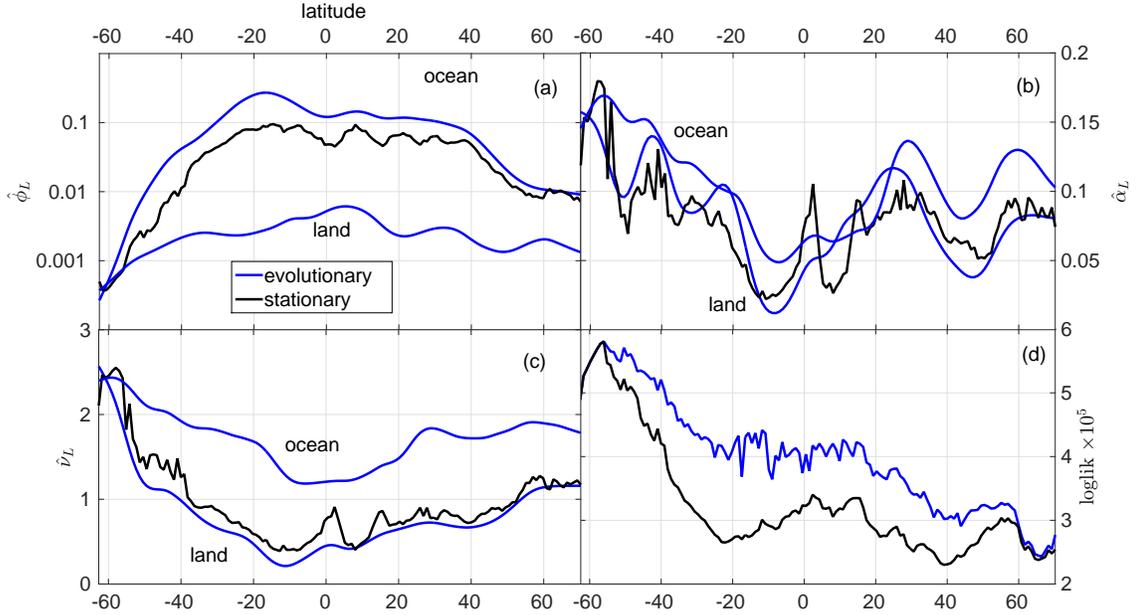}}
\caption{Comparison of the models with evolutionary spectrum \eqref{quax_eq1} and the axially symmetric model with constraint \eqref{bandstat} in terms of (a) $\log(\hat{\phi}_{L_m})$ and $\log(\hat{\phi}^{j}_{L_m})$ for $j=1,2$, (b) $\hat{\alpha}_{L_m}$ and $\hat{\alpha}^j_{L_m}$, (c) $\hat{\nu}_{L_m}$ and $\hat{\nu}^j_{L_m}$, and (d) loglikelihood. A smoothing spline has been applied to the estimated parameters for the evolutionary spectrum approach in a-c since the pattern were less regular.}
\label{single_band}
\end{figure}

The sequential model-fitting procedure can also be used to fit axially symmetric versions of the model for the innovations. This involves imposing the constraint \eqref{bandstat} in approximation 2. In Figure \ref{single_band} we see a comparison of the evolutionary spectrum model with the axially symmetric model (i.e. with constraint \eqref{bandstat}) in terms of estimated parameters and loglikelihood. Figure \ref{single_band}a-c shows how land and ocean parameter estimates for the evolutionary spectrum are very different from those of the stationary model, and how there is a consistent difference across latitude. Figure \ref{single_band}c shows how the smoothness parameter is smaller for land than for ocean which implies, as noticed in Figure \ref{85_norm_per}, that ocean temperatures tend to have a smoother behavior across the same band compared to land. Given the very large size of the data set, the parameter estimates are very precise and the estimated standard deviations\footnote{computed at each step conditional to the previous steps. For example, the standard deviations for $(\hat{\phi}^j_L,\hat{\alpha}^j_L,\hat{\nu}^j_L)$ are estimated conditional on the temporal parameters. } are two orders of magnitude smaller than the point estimates. Thus, we chose not to report the confidence intervals as they are very small compared to the differences across latitudes. We also report in Figure \ref{single_band}d the individual maximum loglikelihoods for each band. The loglikelihood shows a noticeable improvement for the evolutionary spectrum approach, especially in the Southern Hemisphere. In latitudes where there is no land, such as the southernmost bands considered (we removed the Antarctic regions) the evolutionary spectrum and the axially symmetric model are the same and thus have the same loglikelihood. 

Approximation 3 would require estimating $\xi_{L_m}$ and $\tau_{L_m}$ when $-23^{\circ}<L<23^{\circ}$ and a constant value for both parameters outside the tropics, for a total of 50 parameters. Since a likelihood maximization for such number of parameters and with tens of millions of data points is not feasible, we first consider \eqref{ar_nstat} for adjacent bands, obtain their estimates independently for every pair of bands, which we denote as $\hat{\xi}^{(2)}_{L_m}$ and $\hat{\tau}^{(2)}_{L_m}$, and consider these estimates as fixed in approximation 3. (Since every band is involved in two fits, by convention at latitude $L_m$ we plug in the estimates from bands $(L_m,L_{m+1})$). The fitted parameters for \eqref{ar_lat} and \eqref{ar_nstat}, along with $\hat{\xi}^{(2)}_{L_m}$ and $\hat{\tau}^{(2)}_{L_m}$, can be found in Figure S4 in the supplement. The stationary model shows some misfit, especially for $\xi$: this is due to model assuming a constant value across latitude for the coherence, while this is significantly smaller in the southernmost regions and at some tropical latitudes. The parameters of the nonstationary AR(1) model \eqref{ar_nstat} instead are fixed and equal to $\hat{\xi}_{L_m}^{(2)}$ and $\hat{\tau}_{L_m}^{(2)}$ in the tropical regions (by construction) and, while still not capturing nontrivial latitudinal patterns outside the tropics, it results in a larger and more satisfactory estimate for $\xi$.

\section{Model Comparison}\label{sec_comparison}
Table \ref{fitres} shows a comparison among a model that assumes spatial independence (denoted \textit{ind}), the axially symmetric model (denoted \textit{ax}), a model with land/ocean evolutionary spectrum with a stationary AR(1) latitudinal process \eqref{coh} (denoted \textit{ev-st}) and one with a nonstationary latitudinal AR(1) process \eqref{ar_nstat} (denoted \textit{ev-nst}). 

\renewcommand\arraystretch{1.3}
\begin{table}[tbhp]
\caption{{\small Comparison between different models in terms of number of parameters (excluding the temporal ones), computational time (hours), normalized restricted loglikelihood \eqref{reml_me}, and Bayesian Information Criterion \citep{sc78}. }}\label{fitres}
\centering
{\begin{tabular}{|c|c|c|c|c|}
\hline
 Model   & \textsl{ind} & \textit{ax} & \textit{ev-st} & \textit{ev-nst} \\ \cline{1-5} 
\# param   & 0 & 428 & 1138 & 1234 \\ \cline{1-5} 
time  (hours)  & 1.4 & 1.5 & 13.8 & 14.8 \\ \cline{1-5}  
$\Delta$loglik/NMT(R-1)  & -2.87 & -0.61 & -0.0018 & 0 \\ \cline{1-5}  
BIC$\times 10^8$ & -0.1677 & -1.0465 & -1.2832 & -1.2839 \\ \cline{1-5}  
\end{tabular}}
\end{table}
The model assuming independence is clearly the fastest to fit, as once the temporal part is estimated, the full likelihood can be evaluated just once. The axially symmetric model requires spatial parameters, but the computational time is almost equivalent and the improvement both in terms of normalized likelihood and BIC is noticeable. The evolutionary spectrum model requires approximately three times more parameters than the axially symmetric model and a noticeable increase of computational time (mostly because of the 2-band step). The resulting model, however, shows a dramatic improvement; the loglikelihood improves by 0.6 units per observation, and improves  the BIC despite the large increase in the number of parameters. \textit{ev-nst} requires more parameters (the plug-in estimates $\hat{\xi}^{(2)}_{L_m}$ and $\hat{\nu}^{(2)}_{L_m}$ at the equator) and there is small indication of a further improvement in the fit.

\begin{figure}[ht]
\hspace{-0.5em}
\centerline{\includegraphics[width=16cm,keepaspectratio]{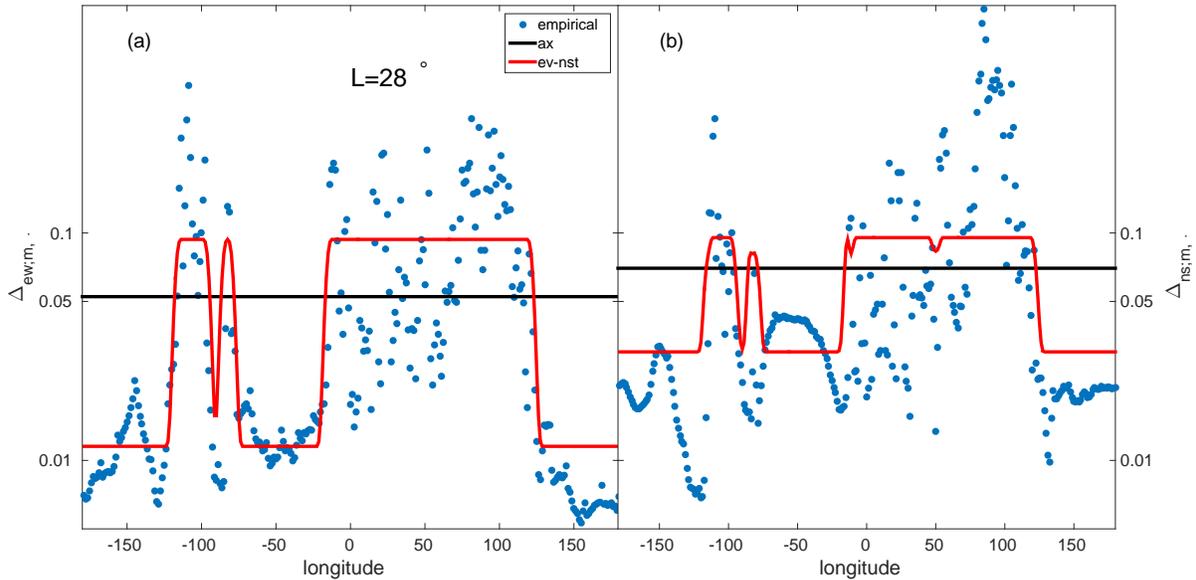}}
\caption{Estimated and fitted variances for the contiguous differences at approximately 28$^{\circ}$ North for different longitudes, averaged across time and realizations. (a): $\Delta_{\text{ew};m,\cdot}$ and (b): $\Delta_{\text{ns};m,\cdot}$ as in \eqref{contr_not}. The vertical axis is plotted on a log scale. }
\label{contr_sband}
\end{figure}

\begin{sloppypar}
To assess the quality of the fit, we compute the contrast variance 
\begin{equation}\label{contr_not}
\begin{array}{lll}
\Delta_{\text{ew};m,n} & = & \frac{1}{KR}\sum_{k=1}^K\sum_{r=1}^R \left\{\widehat{\mbf{H}}_r(L_m,\ell_n,t_k)-\widehat{\mbf{H}}_r(L_m,\ell_{n-1},t_k)\right\}^2,\\[7pt]
\Delta_{\text{ns};m,n} & = & \frac{1}{KR}\sum_{k=1}^K\sum_{r=1}^R \left\{\widehat{\mbf{H}}_r(L_m,\ell_n,t_k)-\widehat{\mbf{H}}_r(L_{m-1},\ell_n,t_k)\right\}^2,
\end{array}
\end{equation}
where ew=east-west and ns=north-south, and compare them with the corresponding fitted quantities according of \textit{ax} (axially symmetric) and \textit{ev-nst} (latitudinally nonstationary evolutionary spectrum). The result for a chosen band at approximately 28$^{\circ}$ North is shown in Figure \ref{contr_sband}. Both panels show the limits of axial symmetry which, assuming longitudinal stationarity, results in a constant value across longitude. This is clearly not adequate for temperature data, as there are significant longitudinal patterns generated by different land/ocean domains; for this latitude, $\Delta_{\text{ew};m,\cdot}$ is nearly ten times larger over land than it is over ocean. The evolutionary spectrum model proposed here is noticeably more flexible and is able to accurately capture the changes across longitude in the contrast variances. It is apparent how different domains have different behaviors, and thus different spatial correlation, and how the fitted \textit{ev-nst} allow for a smoother spatial behavior over the ocean. The evolutionary spectrum proposed is particularly effective in capturing $\Delta_{\text{ew};m,\cdot}$ in Figure \ref{contr_sband}a, while some misfit is present in the Pacific Ocean for the north-south contrast variances in Figure \ref{contr_sband}b. 
\end{sloppypar}

\begin{figure}[ht]
\hspace{-0.5em}
\centerline{\includegraphics[width=16cm,keepaspectratio]{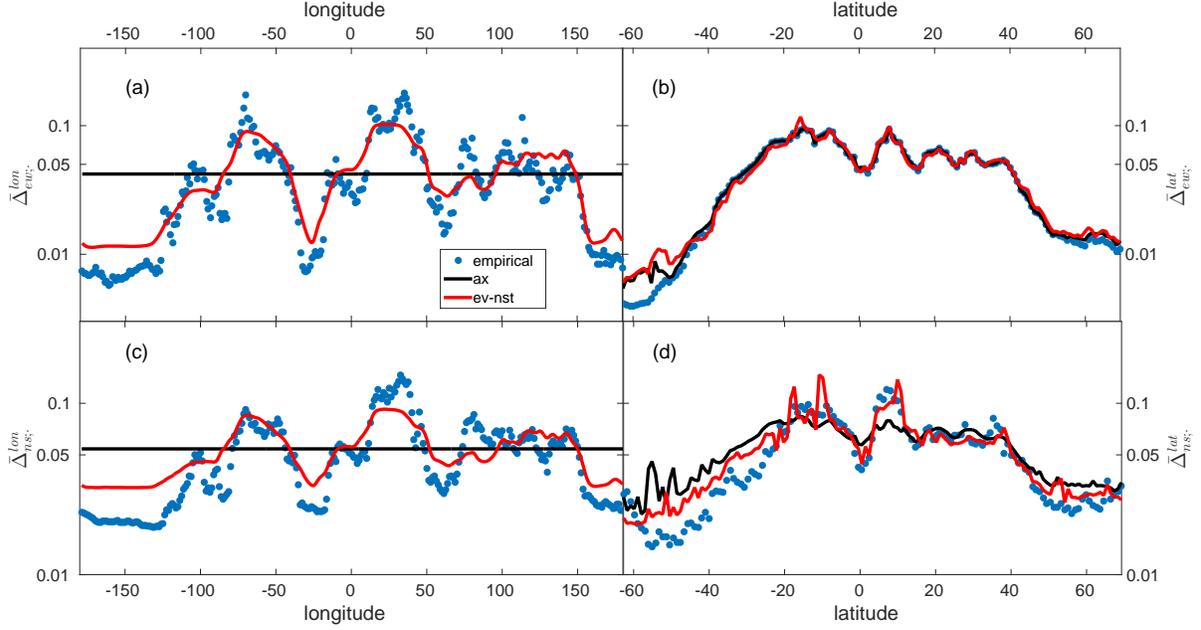}}
\caption{Estimated and fitted averaged contrasts variances as in \eqref{contr_m}. (a): $\bar{\Delta}^{\text{lon}}_{ew;\cdot}$ (b): $\bar{\Delta}^{\text{lat}}_{ew;\cdot}$, (c): $\bar{\Delta}^{\text{lon}}_{ns;\cdot}$ and (d): $\bar{\Delta}^{\text{lat}}_{ns;\cdot}$. The vertical axis is plotted on a log scale.}
\label{contr_means}
\end{figure}

To assess the fit for multiple latitudes, we compute the average contrast variance
\begin{equation}\label{contr_m}
\begin{array}{lll}
\bar{\Delta}^{\text{lat}}_{j;m} & = & \frac{1}{N}\sum_{n=1}^N \Delta_{j;m,n},\\[7pt]
\bar{\Delta}^{\text{lon}}_{j;n} & = & \frac{1}{M}\sum_{m=1}^M \Delta_{j;m,n},
\end{array}
\end{equation}
where $j=\{\text{ew}, \text{ns}\}$. In Figure \ref{contr_means}a-b, the values for $j=\{\text{ew}\}$ contrasts are shown, and while both \textit{ax} and \textit{ev-nst} are able to capture longitudinally averaged variances in panel (b) (apart from a misfit of both models in the southernmost bands), only \textit{ev-nst} is able to capture the pattern in latitudinally averaged variance in (a), since \textit{ax} assumes constant variance across longitudes. Figure \ref{contr_means}c-d shows the values for $j=\{\text{ns}\}$. Similar remarks as in the previous case hold, but \textit{ev-nst} does not fully capture the patterns of latitudinally averaged variance in panel (c), while the two models performs similarly (and with some degree of misfit in the southernmost latitudes) in longitudinally averaged variances in panel (d).

\section{Simulating the initial condition ensemble}\label{sec_simulation}

We now proceed with simulating surrogate (emulated) runs according to the evolutionary spectrum with nonstationarity in latitude \eqref{ar_nstat}. From \eqref{reml_me}, the mean can be estimated as $\hat{\bsy{\mu}}=\mbf{\bar{T}}$. For each location, we fit a cubic polynomial smoothing spline $\widetilde{\mbf{T}}_{m,n}$ from $\lambda \sum_{k=1}^K \{\mbf{\bar{T}}(L_m,\ell_n,t_k)-\widetilde{\mbf{T}}_{m,n}(t_k)\}^2+(1-\lambda)\sum_{k=1}^K \left|\frac{\mathrm{d}^2 \widetilde{\mbf{T}}_{m,n}}{\mathrm{d}k^2}(t_k) \right|^2$ with mild penalty term $\lambda=0.01$ since the climate is slowly varying \citep{ca16}, and we denote by $\widetilde{\mbf{T}}=(\widetilde{\mbf{T}}_{1,1},\ldots,\widetilde{\mbf{T}}_{M,N})$. To generate a simulation, the following steps are required:

\begin{itemize}
\item generate $\mbf{e}_{L_m}(c)\iid \mathcal{N}(0,1-\varphi_{L_m}(c)^2)$ with $\varphi_{L_m}(c)$ as in \eqref{ar_nstat},
\item compute $\widetilde{\mbf{H}}_{L_m}(c)$ with \eqref{ar_lat} and \eqref{ar_nstat},
\item compute $\mbf{H}_{r}(L_m,\ell_n,t_k)$ with \eqref{lo_reg},
\item compute $\bsy{\varepsilon}_r$ with \eqref{AR1},
\item obtain the surrogate run as $\widetilde{\mbf{T}}+\bsy{\varepsilon}_r$.
\end{itemize}

Once the parameters have been estimated, a common laptop can generate hundreds of surrogate runs almost instantaneously with the aforementioned steps.  

\begin{figure}[ht]
\hspace{-0.5em}
\centerline{\includegraphics[width=16cm,keepaspectratio]{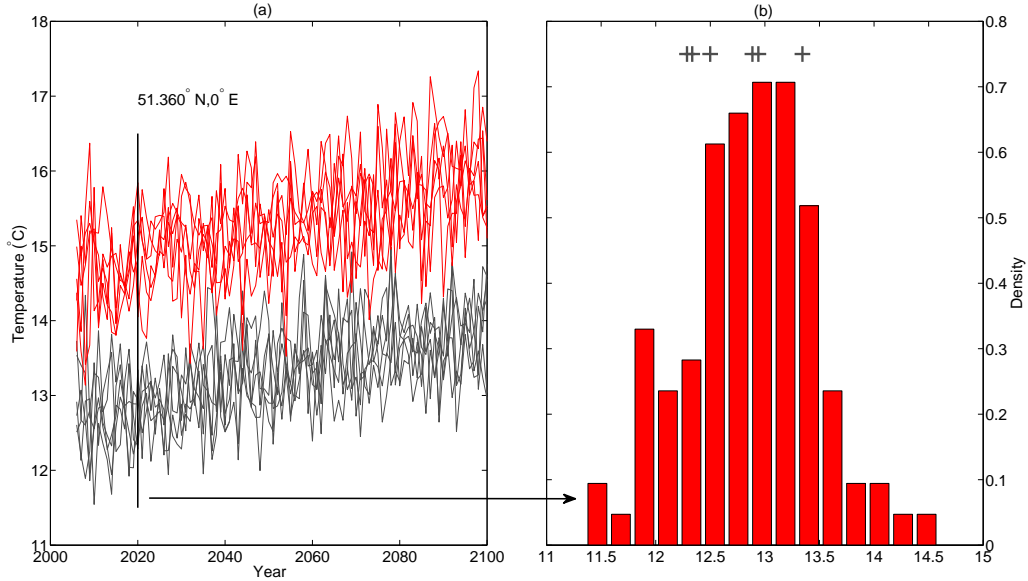}}
\caption{Comparison of climate model output with surrogate runs. (a) The six realizations of the climate model near London (in gray) are shown against six surrogate runs (in red, offset by 2 C$^\circ$). (b) Histogram of the distribution of temperature for the year 2020 for the 100 surrogate runs. The gray crosses above represent the six realizations from the ensemble for the same year.}
\label{compare}
\end{figure}

In Figure \ref{compare} we show a comparison of the six runs from the climate model ensemble and the surrogate runs in terms of temperature series near London. In panel (a), the six climate model runs are compared with six surrogate runs (offset by 2 C$^\circ$ to avoid superimposition). The two groups show the same trend and the same variance, but the statistical model allows to generate more runs, so that it is possible to have a better assessment of the temperature uncertainty at a given year. (In this context, by ``uncertainty'' we mean ``uncertainty due to initial conditions''. We do not consider the uncertainty due to physical parameter calibration or forcing scenario.) In panel (b), we see how having just six climate model runs is poorly informative of the projection uncertainty for 2020, while with 100 surrogate runs it is possible to have a better assessment.

Although a comparison on a single location does not inform about the ability of the statistical model to capture the spatial variability, it is possible to produce animations of surrogate runs to detect if the spatial patterns are qualitatively consistent. \cite{ge15} have discussed in detail how climate model output and statistical surrogates can be compared in the case of three dimensional annual temperatures by using a virtual reality environment. In this work, we produce movies for one climate model run and a surrogate run (both in the supplementary material), which qualitatively shows similar large-scale features.

\section{Conclusion and discussion}\label{sec_conclusion}

In this work we introduced a new class of spectral models that is able to incorporate geographical information to capture the nonstationary behavior of global data across longitude. We further introduced a nonstationary structure across latitude that allows for a more flexible and general description of the dependence among different bands. The evolutionary spectrum model we developed vastly outperforms axially symmetric models, showing improved performance under common model selection metrics and the estimation of the contrast variances. By using appropriate diagnostics, we show how this model is able to capture patterns across longitude that would be constant under the assumption of axial symmetry. The proposed model can be also used to incorporate further geographical information, such as orography, or can be applied to other physical quantities whose dynamics are known to be influenced by large scale geographical features, such as precipitation or winds.

The likelihood of the proposed model can be written in a computationally convenient form, which is almost as fast as in the axially symmetric case and can be successively approximated with a highly parallelizable algorithm while still preserving the main space-time structure, as shown in the diagnostics. While in this work the approximation blocks and the order of approximation (time, longitude, latitude) have been suggested by the particular problem, the multi-step approximation presented can be applied to any large space-time data set where the nature of the problem suggests blocks: for example, in functional Magnetic Resonance Imaging the brain can be naturally divided into regions of interests when monitoring cognitive tasks. The analysis was performed on a state-of-the-art workstation, allowing distributed computing to optimize the efficiency, and achieving a fit in less than one day for more than 20 million data points. This model consists of 1234 spatial parameters and 121824 temporal parameters (three for each location), thus achieving a compression rate of 3:100, which is vastly superior to traditional compression algorithms which can achieve at most a 1:5 rate. The fit requires substantial computational power, but the estimates can then be used to generate surrogate runs almost instantaneously on a laptop. 

Estimating all the parameters at once would require maximizing a likelihood over more than 100,000 parameters for more than 20 million data points, a extremely challenging task to perform within a reasonable time even with the most advanced computational facilities. Thus, we devised a step-wise estimation procedure with plug-in estimates from previous stages, which result in error propagation across stages that needs to be detected and mitigated. Bias propagation can be detected with diagnostic figures such as \ref{contr_sband} and \ref{contr_means}, and can be mitigated with an intermediate 2-band step before the latitudinal modeling to adjust the single band point estimates. Estimation uncertainty propagation in this context is of less concern, as given the considerable size of the data set, the estimated standard deviation is several orders of magnitude smaller than point estimates and the bias largely dominates the error propagation.

Despite the substantial improvements in flexibility, statistics-based compression is intrinsically dependent on the statistical model assumptions. The proposed methodology cannot generate surrogate runs that substitute the climate model, as the complex nonlinear dynamics of annual surface temperatures cannot be fully represented by a Gaussian process. As for emulators, our statistical model is to be regarded as a useful stochastic approximation that could help climate model users to test initial scientific hypotheses, but should not be used to perform a full geophysical investigation.



\baselineskip 21.3 pt
\bibliographystyle{asa}
\bibliography{citations}

\end{document}